\documentclass[aps]{revtex4}
\usepackage{amsfonts}
\usepackage{amsmath}
\usepackage{amssymb}
\newcommand{\be}{\begin{equation}}
\newcommand{\ee}{\end{equation}}
\newcommand{\ben}{\begin{eqnarray}}
\newcommand{\een}{\end{eqnarray}}
\newcommand{\bb}{\bibitem}

\newcommand{\wt}{\widetilde}
\newcommand{\dslash}{\partial\!\!\!/}
\newcommand{\aslash}{A\!\!\!/}

\begin{document}
\title{Dual equivalence in models with higher-order derivatives}
\author{D. Bazeia,$^a$ R. Menezes,$^a$ J.R. Nascimento,$^a$
R.F. Ribeiro,$^a$ and C. Wotzasek$^b$}
\affiliation{{\em$^{b}$Departamento de
F\'\i sica, Universidade Federal da Para\'\i ba, 58051-970 Jo\~ao
Pessoa, PB, Brazil}\\{\em${}^{a}$Instituto de F\'\i sica,
Universidade Federal do Rio de Janeiro, 21945-970 Rio de Janeiro,
RJ, Brazil}}
\date{\today}

\begin{abstract}
We introduce a class of higher-order derivative models in $(2,1)$ space-time
dimensions. The models are described by a vector field, and contain a
Proca-like mass term which prevents gauge invariance. We use the gauge
embedding procedure to generate another class of higher-order derivative
models, gauge-invariant and dual to the former class. We show that the results
are valid in arbitrary $(d,1)$ space-time dimensions when one discards the
Chern-Simons and Chern-Simons-like terms. We also investigate duality
at the quantum level, and we show that it is preserved in the quantum
scenario. Other results include investigations concerning the gauge embedding
approach when the vector field couples with fermionic matter, and when one adds
nonlinearity.
\end{abstract}
\maketitle

%%%%%%%%%%%%%%%%%%%%%%%%%%%%%%%%%%%%%%%%
\section{Introduction}
It is hardly necessary to remind the remarkable and powerful properties
of the duality mapping as an analytical tool in Field Theory as well as
in String Theory \cite{agz}. On the other hand, the interest in the study
of theories involving higher-order derivative is by now well-appreciated
and remains intense. Within the context of the Maxwell theory, generalizations
involving higher-order derivatives can be found in \cite{bi,pr,po}. More
recently,
in Refs.~{\cite{dj,dn}} one finds generalizations that involve both the
Maxwell and Chern-Simons (CS) terms \cite{cs1,cs2,cs3}. In the present work
we study duality symmetry in extended theories, which contain higher-order
derivatives involving both the Maxwell and CS terms.

The interest in the subject has been recently fueled by several
motivations, in particular by issues related to String Theory. As one knows,
string theories engender the feature of containing interactions
that may involve an infinite number of space-time derivatives, and that may
lead to standard Field Theory in the low energy limit. Thus, higher-order
derivative contributions would certainly appear when one consider
next-to-leading orders in the energy. Based upon such
interesting possibilities, there is renewed interest in investigating
higher-order derivative models, including the case of tachyons
\cite{w,ks,t1,t2,t3,t4,t5,t6}. Other lines of investigations can be found
for instance in \cite{wes,hli}, and also in gravity, where higher orders
in the scalar curvature $R$ are considered in the so-called nonlinear gravity
theories -- see, for instance, the recent works \cite{pmt,muk} and references
therein. 

The presence of higher-order derivatives to control the behavior of systems
is not peculiar to String and Field theory. It may also appear in condensed
matter, and can be important to describe higher-order phase
transitions \cite{pip}, as for instance the case of the fourth-order
transition in superconducting ${\rm Ba}_{0.6}{\rm K}_{0.4}{\rm BiO}_3$,
which is described in \cite{khg}. Furthermore, sometimes one has to include
higher-order derivative of the order parameter to correctly describe pattern
formation in chemical reactions, and in other branches of nonlinear
science \cite{kur,wal}.

We investigate the subject dealing with issues that appear very
naturally in Field Theory. Specifically, we examine the duality
mapping of higher-derivative extensions of self-dual (SD) and
Maxwell-Chern-Simons (MCS) theories, including the presence of
fermionic matter. We work in the $(2,1)$ dimensional space-time,
with $\varepsilon^{012}=\varepsilon_{012}=1$; our metric tensor
has signature $(+,-,-)$ and we use natural units. We start in
the next Sec.~{\ref{sec:exe}}, where we examine
the dual mapping of the MCS-Proca model with the MCS-Podolsky
theory. Our methodology makes use of the gauge embedding
procedure, an approach which has been shown to work very
efficiently to unveil the dual partner of a specific model
\cite{e1,e2,e3,e4,e5}. The investigation of the MCS-Proca model is
new, and we use it to set stage to generalize the model to
higher-order derivatives. We develop this generalization in
Sec.~{\ref{sec:gen}}, where we introduce the main model,
which is defined in terms of $n=1,2,...$, and in Sec.~{\ref{sec:og}},
where we consider the case $n\to\infty$. We show in Sec.~{\ref{sec:gen}}
that if one discards the CS and CS-like terms, the results are then valid
in arbitrary $(d,1)$ space-time dimensions. We also show in
Sec.~{\ref{sec:og}} that the duality is preserved at the quantum level.
In Sec.~{\ref{sec:mat}} we examine the presence of matter, coupling fermions
to the system. In Sec.~{\ref{sec:nli}} we change the model introduced in
Sec.~{\ref{sec:gen}} to include the presence of nonlinearity.
We end our work in Sec.~{\ref{sec:cc}}, where we present our
comments and conclusions.

%%%%%%%%%%%%%%%%%%%%%%%%%%%%%%%%%%%%%%%%%%%%%%%%%%%%%%%%%
\section{Duality transformation in the MCS-Proca model}
\label{sec:exe}

It is well know that both the SD and the MCS models are dual
representations of the same dynamics: they carry one massive
degree of freedom of definite helicity determined by the relative
sign of CS term. However, the SD representation hides a gauge symmetry,
which is explicit in the MCS model. This is easily seen when
one establishes the correspondence
$f_\mu\to F_\mu\sim\varepsilon_{\mu\nu\rho}\partial^\nu A^\rho$,
which maps the SD field $f_\mu$ into the dual of the basic field
$A_\mu$ of the MCS model. We notice that in the above dual mapping,
one relates a gauge non-invariant model with an equivalent, gauge
invariant theory. In this process, however, the non-gauge field is
identified with a special form containing the derivative of the gauge
field of the dual model. The identification of this mechanism is crucial
for the generalizations that we will implement below.

In this section we elaborate on a prototype of the study we intend  develop
in this paper. We discuss the theoretical motivations for studying higher
derivatives dualities, present a physical scenario for the applications
of the ideas elaborated and the quantum implications.

In the past, there have appeared several different ways
of extending electrodynamics, trying to smooth infrared or
ultraviolet singularities that appear at large or short distances.
One is the Born-Infeld \cite{bi} type of generalization, and others
include the generalizations introduced by Proca \cite{pr} and
by Podolsky \cite{po}. The Born-Infeld approach involves nonlinear
extension, which will be considered in Sec.~{\ref{sec:nli}}.
The Proca model involves the addition of a mass term for the vector
field, which was introduced to smooth infrared singularities.
The Podolsky model involves higher-order derivatives and was
introduced to smooth ultraviolet singularities. Thus, the Proca
and Podolsky model deal with dual aspects of the electromagnetic
interaction, and so they may be connected by some duality procedure.

The dual aspects of electrodynamics that appears in the Proca and Podolsky
models constitute the central subject of the present Section.
This duality is also of importance in the extension of the bosonization
program from D=2 to higher dimensions. The mechanisms of the 3D bosonization
in particular are quite dependent on the duality results involving the
presence of the Chern-Simons term. Besides providing us with the proper
scenario for the applications of our ideas, bosonization will also be
crucial for the interpretation of the new parameters in the Podolsky
extension.

%%%%%%%%%%%%%%%%%%%%%%%%%%%%%%%%%%%%%%%%%%%%%%%%%
\subsection{The duality Procedure}
\label{sec:dual}

In order to exemplify the general procedure of duality, we consider
the more general model
 \be\label{mcsp}
{\cal L}=\frac{m^2}{2} A^\mu A_\mu-\frac{a}{4}F^{\mu\nu}F_{\mu\nu}
-\frac{1}{2}m\varepsilon_{\mu\nu\lambda}A^\mu\partial^\nu A^\lambda
 \ee
where $m$ is the mass parameter. We have introduced the real and dimensionless
parameter $a$ in order to obtain the SD model for $a=0$, or the MCS-Proca
model for $a=1$.

The equation of motion involves second-order derivatives, and
the model is supposedly dual to a generalized MCS-Podolsky theory. To verify
this assumption we follow the gauge embedding procedure \cite{e1,e2,e3,e4,e5}
to construct its dual equivalent model. Firstly, we compute the Euler vector
associated with the MCS-Proca theory. We get
 \be\label{eq3}
K_\mu = m^2 A_\mu-m\,\varepsilon_{\mu\nu\lambda}
\partial^\nu A^\lambda+a\,\partial^\nu F_{\nu\mu}
 \ee
The first iteration leads to
 \be
{\cal L}_1={\cal L}_0-K^\mu B_\mu
 \ee
where ${\cal L}_0$ is identified with MCS-Proca Lagrangian given by
Eq.~(\ref{mcsp}). Also, $B_{\mu}$ is an auxiliary field, which varies
according to
 \be\label{eq5}
\delta B_\mu =\delta A_\mu =\partial_\mu \Lambda
 \ee
This choice makes the non-invariant term in ${\cal L}_0 $ to cancel
with the term $K_\mu\delta B^\mu$. Therefore
 \be
\delta{\cal L}_1=-B_\mu\delta K^{\mu}=-\frac{m^2}{2}\delta(B^2)
 \ee
Thus, we can write the gauge invariant second-iterated Lagrangian
 \be
{\cal L}_2={\cal L}_0-K^\mu B_\mu+\frac{m^2}{2} B^2
 \ee
This ends the iteration process, and we can eliminate the
auxiliary field to obtain the dual model
 \be
{\cal L}_{D}={\cal L}_0-\frac{K^2}{2m^2}
 \ee
or better
 \be\label{mcspo}
{\cal L}_{D}=\frac{a-1}{4}F^{\mu\nu}F_{\mu\nu}
+\frac{1}{2}m\varepsilon_{\mu\nu\lambda}A^\mu\partial^\nu A^\lambda-
\frac{a^2}{2m^2}\partial_\mu F^{\mu\nu}\partial^\lambda
F_{\lambda\nu}+\frac{a}{m}\varepsilon^{\mu\nu\lambda}\partial_\nu
A_\lambda\partial^\rho F_{\rho\mu}
 \ee
This is the generalized MCS-Podolsky model -- see Refs.{\cite{po,dj,dn}} for
further informations on this and other, related models.
We recall that the Podolsky model was introduced in order to smooth
ultraviolet singularities. In this sense, we notice that our approach
for duality is working standardly, since we are linking infrared and
ultraviolet problems, in MCS-Proca and MCS-Podolsky models.

In the above investigation,
if $a=0$ the MCS-Proca model in (\ref{mcsp}) becomes
the standard SD model; in this case, in the generalized MCS-Podolsky
model (\ref{mcspo}) one kills the Podolsky terms, and we are led back
to the MCS model. On the other hand if $a=1$ we get to an extended
CS-Podolsky model. 

It seems important, at this juncture, to establish the scenario
at which the parameter $a$ becomes a physical quantity. This is done next
where we show that action (\ref{mcspo}) for the generalized MCS-Podolsky model
is the low energy effective action for a self-interaction fermionic model. 
To set the problem in a proper perspective, we review shortly the
bosonization procedure.

%%%%%%%%%%%%%%%%%%%%%%%%%%%%%%%%%%%%%%%%%%%%%%%%%%%%%
\subsection{Physical Interpretation}
\label{sec:bos}

It is well known by now that the bosonization in D=3  mapps a massive scalar
particle coupled to a Chern-Simons gauge field into a massive Dirac fermion
for a special value of the Chern-Simons coupling. This is a relevant issue
in the context of transmutation of spin and statistics with interesting
applications to problems both in quantum field theory and condensed matter
physics. This boson-fermion transmutation is a property which holds only
at very long distances, namely, at scales long compared with the Compton
wavelength of the particle. As so these results hold to lowest order in an
expansion in powers of the inverse mass of the particle.

The equivalence of the three dimensional effective electromagnetic action
of the $CP^1$ model with a charged massive fermion to lowest order in
inverse (fermion) mass has been proposed by Deser and Redlich \cite{DR}.
Using the results of \cite{DR}, bosonization was extended from two to three
dimensions \cite{FS}. However, contrary to the two-dimensional case where the
fermionic determinant can be exactly computed, bosonization in higher
dimensions is not exact and, in the general case, it has a non-local structure.
However, for the large mass limit in the one-loop of perturbative evaluation,
a local expression materializes.  Indeed, it has been established \cite{FS},
to the leading order in the inverse fermionic mass, an identity between the
partition functions for the three-dimensional Thirring model and the
topologically massive U(1) gauge theory, whose dynamics is controlled by a
Maxwell-Chern-Simons action. Here we show that the contribution
next-to-leading order leads to the MCS-Podolsky model via duality
transformation.

Below we compute the low energy sector
of a theory of massive self-interacting fermions, the massive Thirring Model 
in 2 + 1 dimensions, that can be bosonized  into a gauge theory, the
Maxwell-Chern-Simons gauge theory and its possible higher derivative
extensions. We start from the fermionic partition function for the
three-dimensional massive Thirring model, following closely the
spirit of \cite{FS},
\be
\label{BB10}
{\cal Z}_{Th}=\int\,{\cal{D}}\bar\psi{\cal{D}}\psi\;e^{-\int\left(
\bar\psi^i (\dslash + M) \psi^i -\frac{g^2}{2N}j^{\mu}j_{\mu}
\right) d^3x}
\ee
with the coupling constant $g^2$ having dimensions of inverse mass.
Here $\psi^i$ are N two-component Dirac spinors and $j^{\mu}$ is a global U(1)
current defined as,
\be
\label{BB20}
j^{\mu}=\bar\psi^i\gamma^{\mu}\psi^i.
\ee
Notice that we have reverted to Euclidean metric in this subsection.
As usual, we  eliminate the quartic interaction by performing a (functional)
Legendre transformation through the identity
\be
\label{BB30}
e^{\int \frac{g^2}{2N}j^{\mu}j_{\mu} d^3x}
=\int {\cal D} A_{\mu}e^{-\int \left(\frac{1}{2}A^{\mu}A_{\mu}+
\frac{g}{\sqrt{N}}j^{\mu}A_{\mu}\right)d^3x}
\ee
after a scaling $A_\mu\to g A_\mu/{\sqrt{N}}$. The partition function
then becomes
\be
\label{BB40}
{\cal Z}_{Th} =  \int {\cal{D}}\bar\psi {\cal{D}}\psi {\cal{D}} A_{\mu}\; e^{
-\int\left(\bar\psi^i (\dslash + M + \frac{g}{\sqrt{N}}\aslash )\psi^i
+\frac{1}{2} A^{\mu}A_{\mu}\right)d^3x}.
\ee
Formally, the fermionic path-integral gives the Dirac operator determinant,
\be
\label{BB50}
\int {\cal{D}}\bar\psi {\cal{D}}\psi e^{-\int\bar\psi^i \left(\dslash + M +
\frac{g}{\sqrt{N}}\aslash\right)\psi^i d^3x} = \det \left(\dslash+ M +
\frac{g}{\sqrt{N}}\aslash\right)
\ee
The determinant of the Dirac operator is an unbounded operator and requires
regularization. Bosonization will depend on the actual computation of this
deteminant, namely, whether it can be computed exactly in a closed form or an
approximate recipe must be enforced. In general it leads to non-local
structures but, under some approximation scheme (like the inverse mass),
a local result emerges. 

This determinant can be computed exactly for $D=2$, both for abelian
and non-abelian symmetries. For the D=3 case this determinant has been
computed in \cite{DR} as an expansion in inverse powers of the fermion mass
giving, in the leading order, the Chern-Simons parity violating term, as well
as parity conserving Maxwell term, which is central to our discussion here,
\be
\label{BB60}
\ln\det\left(\dslash+M+\frac{g}{\sqrt{N}}\aslash\right)=\pm\frac{ig^2}{16\pi}
\int\epsilon_{\mu\nu\alpha} F^{\mu\nu} A^{\alpha} d^3x -
\frac{g^2}{24\pi M}\int d^3x  F^{\mu\nu} F_{\mu\nu} + O(\partial^2/M^2)
\ee
We bring these results into the partition function to get
\be
\label{BB80}
{\cal Z}_{Th} = \int {\cal{D}} A_{\mu}  e^{-{\cal S}_{eff}[A_{\mu}]}
\ee
where ${\cal S}_{eff}[A_{\mu}]$ is given, up to order $1/m$, by 
\be
\label{BB90}
{\cal S}_{eff}[A_{\mu}] = \frac{1}{2}\int d^3x \left(A_{\mu}A^{\mu} \mp
\frac{ig^2}{4\pi}\epsilon^{\mu\alpha\nu}A_{\mu}\partial_{\alpha}A_{\nu}-
\frac{g^2}{12\pi  M}  F^{\mu\nu} F_{\mu\nu} \right)
\ee
which, after the scaling $A_\mu \to m A_\mu$, with $m = {4\pi}/g^2$. 
With these identifications we find $a=\frac{2m}{3M}$. In conclusion,
we have established the following identification:
\be
\label{BB100}
{\cal Z}_{Th} \approx {\cal Z}_{MCS-Proca}
\ee
which is valid to next-to-leading order in $1/M$.

In the preceeding section we have established the dynamical equivalence
between the model defined by ${\cal S}_{MCS-Proca}$ the MCS-Podolsky theory.
This proves the equivalence, to this order in $1/M$ expansion of the partition
functions for  Thirring model and  the MCS-Podolsky theory:
\be
\label{BB120}
{\cal Z}_{Th} \approx {\cal Z}_{MCS-Podolsky}
\ee
It expresses the equivalence between a theory of $3$-dimensional interacting
fermionic theory and Maxwell-Chern-Simons vectorial bosons in the long
wavelength aproximation. It is interesting to observe that the Thirring
coupling constant $g^2/N$ in the fermionic model is mapped into the inverse
mass spin $1$ massive excitation. In the same fashion, we may identify each
term of the generalization to higher-order derivatives studied below, with
terms following from the  $1/M$ expansion of the fermionic determinant.
This is however beyond the scope of the present work.

%%%%%%%%%%%%%%%%%%%%%%%%%%%%%%%%%%%%%%%%%%%%%%%%%%%%%
\section{Generalization to higher-order derivatives}
\label{sec:gen}

 In this Section we use the duality procedure which connects
the SD and MCS models, and the MCS-Proca and MCS-Podolsky models, in order
to extend the formalism to generalized models involving higher-order
derivatives. Evidently, the presence of higher-order derivatives introduces
longer distances effects and is of interest to String Theory, and to
investigations involving phase transition. We split the subject
into two subsections, the first exploring the relevant classical issues,
and the second dealing with duality at the quantum level.

%%%%%%%%%%%%%%%%%%%%%%%%%%%%%%%%%%%%%
\subsection{Duality at the clasical level}

Let us rescale fields and coordinates as
$A\to m^{1/2} {\bar A}$ and $x\to m^{-1}{\bar x}$, in order to work
with dimensionless quantities. We rewrite ${\bar A}$ and ${\bar x}$
as $A$ and $x$ again, and we define the general field $A^{(n)}_\mu$
through the recursive relation
 \be\label{gf}
A^{(n)\mu}\equiv\epsilon^{\mu\nu\rho}\partial_\nu
A_\rho^{(n-1)}
 \ee
Here we use $A^{(0)}_\mu=A_\mu$ to represent the basic field. The subscript
$n$ which identifies the field also shows the number of derivatives one has
to perform in the basic field. We notice that
in the above relation (\ref{gf}) the field $A^{(n)}_\mu$ is related to
the field $A^{(n-1)}_\mu$ by means of a derivative, which maintains
the mechanism we have identified in the former Sec.~{\ref{sec:exe}},
where the dual relation between the fields involves a derivative.

For this field we now define the general field strength,
 \be\label{FG}
F^{(n)}_{\mu\nu}\equiv\partial_\mu A^{(n)}_\nu-\partial_\nu
A^{(n)}_\mu.
 \ee
With this, the Maxwell term is proportional to
$A^{(1)}_\mu A{^{(1)\mu}}$, and can be generalized to
$A^{(n)}_\mu A^{(n)\mu}$.

We define the antisymmetric tensor $G^{(n)}_{\mu\nu}$ such that
$G^{(0)}_{\mu\nu}=F_{\mu\nu}$, and
 \be\label{G}
G^{(n)}_{\mu\nu}\equiv\partial_{[\mu}\partial^\lambda
G^{(n-1)}_{\lambda\nu]}=\partial_{\mu}\partial^\lambda G^{(n-1)}_{\lambda
\nu}-\partial_{\nu}\partial^\lambda G^{(n-1)}_{\lambda\mu}
 \ee
which is important to relate $A^{(n)}_\mu$ to $A^{(0)}_\mu$. We
notice that $G^{(n)}_{\mu\nu}$ contains $2n+1$ implicit
derivatives in $A_{\mu}$. So, the general field is, for $n$
positive,
 \be\label{GG}
A^{(n)}_\mu=\begin{cases}
(-1)^{\frac{n}{2}}\partial^\nu\;
G^{(n/2-1)}_{\nu\mu}&{\rm for}\;n\;{\rm even}
\\ \\\frac{1}{2}(-1)^{\frac{n-1}{2}}\varepsilon_{\mu}^{\;\;\nu\lambda}\;
G^{(n/2-1/2)}_{\nu\lambda}&{\rm for}\;n\;{\rm odd}
 \end{cases}
 \ee

To include CS-like terms into the proposed generalization we first
notice that the CS term can be written as
$\varepsilon^{\mu\nu\lambda}A_\mu\partial_\nu
A_\lambda=A_\mu^{(0)}A{^{(1)\mu}}$. Thus, it induces the general form,
$A_\mu^{(i)}A^{(j)\mu}$, where $i,j$ are non-negative integers.
We notice that if $i=j$ we obtain the generalized Maxwell term.
Thus, we can modify the generalized Maxwell model appropriately,
to include extended CS contributions. Furthermore,
we can prove the identity
 \be\label{AA}
A_\mu^{(i)}A^{(j)\mu}=A_\mu^{(i-1)}A^{(j+1)\mu}+
\varepsilon^{\mu\nu\lambda}\partial_\mu
\left(A^{(i-1)}_\nu A^{(j)}_\lambda\right)
 \ee
The last term is a total derivative, which can be discarded
since it does not change the action when this identity is used in the
Lagrange density of the corresponding model.

We see that for $i+j$ even we can write, apart from a total
derivative,
 \be
A_\mu^{(i)}A^{(j)\mu}=A_\mu^{\left(\frac{i+j}{2}\right)}
A^{^{\left(\frac{i+j}{2}\right)}\mu}
 \ee
As a result, in the action we do not need to split $A_\mu^{(i)}A^{(j)\mu}$
into generalized Maxwell and extended CS terms.
We introduce a single term, characterized by the sum $i+j$:
for $i+j$ even we get a generalized Maxwell term, and for $i+j$
odd we obtain an extended CS term. For example,
$A_\mu^{(0)}A^{(1)\mu}$ and $A_\mu^{(0)}A^{(2)\mu}$ reproduce
the standard CS and Maxwell terms, respectively.

We use the above results to introduce the model
 \be\label{gt}
{\cal L}^{(n)}=\sum^n_{i=0}{r_i}
A_\mu^{(0)}A^{(i)\mu}
 \ee
where $n$ is integer, and $r_i$ are real and dimensionless parameters,
with $r_0\neq0$. The equation of motion is
 \be\label{e1}
\sum_{i=0}^{n}r_i A_\mu^{(i)}=0
 \ee
This equation allows writing $\partial^\mu A_\mu^{(0)}=0$, which shows that
no longitudinal mode propagates in the theory. Besides, we can use this
condition to rewrite Eq.~(\ref{GG}) in the simpler form
\be
A^{(n)}_\mu=\begin{cases}
(-1)^{\frac{n}{2}}\Box^{\frac{n}{2}}A^{(0)}_\mu &{\rm for}\;n\;{\rm even}
\\ \\
(-1)^{\frac{n-1}{2}}\Box^{\frac{n-1}{2}}\varepsilon_{\mu\nu\lambda}
\partial^{\nu}A^{(0)\lambda} &{\rm for}\;n\;{\rm odd}
\end{cases}
\ee

With the aim to build an Abelian gauge model, we consider
the variation
$\delta A_\mu=\delta A^{(0)}_\mu=\partial_\mu\Lambda$; as before,
$\Lambda$ is a local infinitesimal parameter. We use this and the definition
for $A^{(n)}_\mu$ to obtain
 \be
\delta A_\mu^{(i)}=0,\;\; i\neq0
 \ee
We vary the Lagrange density in (\ref{gt}) to obtain
 \be
\delta{\cal{L}}^{(n)}=K^\mu\delta A_\mu^{(0)}
 \ee
where the local Noether current is defined as
 \be\label{GK}
K_\mu \equiv 2 \sum_{i=0}^{n} r_i A_\mu^{(i)}
 \ee

We introduce an auxiliary vector field $a_\mu$, and we couple it
linearly to the Euler vector
 \be
{\cal L}^{(n)}_1={\cal L}^{(n)}-a^\mu K_\mu
 \ee
We chose $\delta a_\mu=\partial_\mu{\Lambda}=\delta A_\mu$. Thus
 \be
\delta{\cal L}^{(n)}_1=-a_\mu\delta K^\mu
 \ee
We consider
 \be
{\cal L}^{(n)}_2={\cal L}^{(n)}_1+r_0 a^\mu a_\mu
 \ee
The procedure ends with the elimination of the auxiliary field $a_\mu$. We
get to
 \be
{\cal L}^{(n)}_{D}={\cal L}^{(n)}-\frac{1}{4r_0}K^\mu K_\mu
 \ee
which is the Lagrange density of the dual model. We use the Euler vector
to get
 \be\label{leff}
{\cal L}^{(n)}_{D}=\sum^n_{i=0}r_i A_\mu^{(0)}A^{(i)\mu}
-\frac1{r_0}\sum^n_{i=0}\sum^n_{j=0}r_ir_jA_\mu^{(i)}A^{(j)\mu}.
 \ee
or better
 \be\label{leff2}
{\cal L}^{(n)}_{D}=-\frac1{r_0}\sum^n_{i=1}
\sum^n_{j=0}r_ir_jA_\mu^{(i)}A^{(j)\mu}.
 \ee
which gives the dual model. We notice that the respective action
is gauge-invariant. This result shows that the model (\ref{gt}) engenders
hidden gauge invariance, in a way similar to the SD model \cite{cs1}.
We can rewrite this result in the form
\be\label{dgt}
{\cal L}_D=\sum^{2n}_{i=1}r'_iA^{(0)}_\mu
A^{(i)\mu}
\ee
where we have used
\be
r_i' = \begin{cases}
-\dfrac{1}{r_0}\displaystyle\sum_{l=1}^i r_l \,r_{i-l}&
\text{for $1\leqslant i \leqslant n$}\\ \\
-\dfrac{1}{r_0}\displaystyle\sum_{l=i-n}^{n} r_l \,r_{i-l} &
\text{for $n<i\leqslant2n$}\end{cases}
\ee

Before ending this Section, let us comment on issues related to the
above duality investigation. The equation of motion that follows from
the dual model (\ref{dgt}) is
 \be\label{e2}
\sum^{2n}_{i=1}r'_i A^{(i)}_\mu=0
 \ee
In Eq.~(\ref{e1}) we eliminate $A_\mu^{(0)}$ to see that these two equations
are identical, thus confirming the dual equivalence between the two theories.

Another issue concerns finding a master theory. To get to this, let us
introduce another field $B^{(i)}_\mu$ such that $B^{(0)}_\mu=B_{\mu}$.
We use $A^{(i)}_\mu$ and $B^{(i)}_\mu$ to write
 \be\label{mas}
{\cal L}^{(n)}_M=r_0 B^{(0)}_\mu B^{(0)\mu}+
2\sum^n_{i=1}r_iB^{(0)}_\mu A^{(i)\mu}-
\sum^n_{i=1}r_i A^{(0)}_\mu A^{(i)\mu}
 \ee
We vary the
corresponding action with respect to $B^{(0)}$ to get
 \be
B^{(0)}=-\frac1{r_0}\sum^n_{i=1}r_i A^{(i)\mu}
 \ee
We use this into (\ref{mas}) to obtain
 \be
{\cal L}^{(n)}_{D}=-\frac1{r_0}\sum^n_{i=1}
\sum^n_{j=0}r_ir_jA_\mu^{(i)}A^{(j)\mu}.
 \ee
which is Eq.~(\ref{leff2}). If we vary the master
action with respect to $A^{(0)}_\mu$ to get
 \be
\sum^n_{i=1}r_i B^{(i)}_\mu=\sum^n_{i=1}r_i A^{(i)}_\mu
 \ee
This result allows writing
 \be
{\cal L}=r_0 B^{(0)}_\mu B^{(0)\mu}+\sum^n_{i=1}r_i B^{(0)}_\mu B^{(i)\mu}
=\sum^n_{i=0}r_i B^{(0)}_\mu B^{(i)\mu}
 \ee
which reproduces Eq.~(\ref{gt}), confirming that
Eq.~(\ref{mas}) is a master or parent theory.

The model that we have introduced is defined by
Eq.~(\ref{gt}); it is controlled by the integer $n$ which we
choose to be $n=1,2,3...$. The simplest case is $n=1$, which
reproduces the SD model. The next case is $n=2$, which gives the
MCS-Proca model investigated in the former Sec.~{\ref{sec:exe}}.
The case $n=3$ gives
 \ben\label{3}
{\cal L}^{(3)}&=&r_0 A^{(0)}_\mu A^{(0)\mu}+
r_1 A^{(0)}_\mu A^{(1)\mu}+r_2 A^{(0)}_{\mu}A^{(2)\mu}+
r_{3}A^{(0)}_{\mu}A^{(3)\mu}\nonumber
\\
&=&r_0 A_\mu A^\mu+r_1\varepsilon^{\mu\nu\lambda}
A_{\mu}\partial_\nu A_\lambda+\frac12r_2 F_{\mu\nu}F^{\mu\nu}+r_3
\varepsilon^{\mu\nu\lambda}A_\mu\partial_\nu \partial^{\rho}F_{\lambda\rho}
 \een
and the other cases follow standardly. For $n=3$ the dual model is
 \be
{\cal L}^{(3)}_{D}=-\frac1{r_0}\sum_{i=1}^{3}\sum_{j=0}^{3}r_ir_j
A^{(i)}_\mu A^{(j)\mu}
 \ee
We use Eqs.~(\ref{G}) and (\ref{GG}) to write
 \ben\label{3d}
{\cal L}^{(3)}_{D}&=&-\frac{r_1}{2}\varepsilon^{\mu\nu\lambda}
A_\mu F_{\nu\lambda}-\frac12\left(r_2+\frac{r_1^2}{r_0}\right)
F_{\mu\nu}F^{\mu\nu}+
\frac12\left(r_3+2\frac{r_1r_2}{r_0}\right)\varepsilon^{\mu\nu\lambda}
F_{\mu\nu}\partial^{\rho}F_{\rho\lambda}-\nonumber
\\
&&\frac1{r_0}(r_2^2+2r_1r_3)\partial_\mu
F^{\mu\nu}\partial^{\lambda}F_{\lambda\nu}-2\frac{r_2r_3}{r_0}
\varepsilon^{\mu\nu\lambda}\partial^\rho F_{\rho\mu}\partial_\nu
\partial^{\alpha}F_{\alpha\lambda}-2\frac{r_3^2}{r_0}\partial^\mu
\partial^\nu F_{\nu\lambda}{\partial_\mu\partial_\rho}F^{\rho\lambda}
 \een
In this example, we can choose the parameters $r_0\neq0,r_1,r_2,r_3$ to find
the dual theory to every model, up to the order $n=3$. For instance, if we
use $r_0=1/2,r_1=-1/2,r_2=-a/2$, and $r_3=0$, and if we re-introduce
dimensional units, we obtain from (\ref{3}) and (\ref{3d}) the former
expressions (\ref{mcsp}) and (\ref{mcspo}) that we have investigated
in Sec.~{\ref{sec:exe}}.

We notice that the presence of the CS and CS-like terms imposes the
restriction, that the Minkowsky space must have $(2,1)$ spacetime dimensions.
Thus, if we set $r_i=0$ for $i$ odd, we eliminate all the CS and CS-like terms,
and our results are then valid in arbitrary $(d,1)$ space-time dimensions.

%%%%%%%%%%%%%%%%%%%%%%%%%%%%%%%%%%%%%%%
\subsection{Quantum duality}

We now explore the issue of quantum duality \cite{dst,go,brr}
for the models examined previously. We start with the Lagrange density for
the master theory, given by Eq.~(\ref{mas}). We extend the model to write
\be\label{masj}
{\cal L}^{(n)}_M(J,j)=r_0B^{(0)}_\mu B^{(0)\mu}+2B^{(0)}_\mu F^\mu(A)-
A_\mu^{(0)}F^\mu(A)+\alpha r_0 B^{(0)}_\mu j^{(0)\mu}+\alpha F^\mu(A)J^{(0)\mu}
\ee
where we have set
\ben
F_\mu(A)=\sum_{i=1}^{n}r_i A_\mu^{(i)}
\\
F_\mu(B)=\sum_{i=1}^{n}r_i B_\mu^{(i)}
\een
to ease calculation. We see that the master theory contains two fields,
thus we have added the external currents $j_\mu$ and $J_\mu$; we see
that $j_\mu$ couples with $B_\mu$, while $J_\mu$ couples
with $A_\mu$ indirectly, through $F_\mu(A)$.

We use (\ref{masj}) to eliminate $A_\mu$. We obtain
$F_\mu(A)=F_\mu(B)+(\alpha/2)F_\mu(J),$ which we use to write
\be
{\cal L}^{(n)}(J,j)={\cal L}^{(n)}+\alpha B^{(0)}_\mu(r_0j^{(0)\mu}+F^\mu(J))
+\frac{\alpha^2}{4}J^{(0)}_\mu F^\mu(J)
\ee
where ${\cal L}^{(n)}$ is the model given by Eq.~(\ref{gt}).
We also use (\ref{masj}) to eliminate $B_\mu^{(0)}$. We get
$B_\mu^{(0)}=-(1/r_0)F_\mu^{(0)}(A)-(\alpha/2)j_\mu^{(0)},$ which we use
to abtain
\be
{\cal L}^{(n)}_D(J,j)={\cal L}^{(n)}_D+\alpha F_\mu(A)(J^{(0)\mu}-j^{(0)\mu})
-\frac{\alpha^2r_0}{4}j_\mu j^\mu
\ee
where ${\cal L}^{(n)}_D$ is the dual model, given by Eq.~(\ref{leff2}).

We define the functional generators $Z^{(n)}_M(J,j),$ $Z^{(n)}(J,j)$
and $Z^{(n)}_D(J,j)$ in the usual way. They allow obtaining
\be
\frac{1}{Z^{(n)}_M}\frac{\delta^2 Z^{(n)}_M}{\delta j_\mu(x)\delta j_\nu(y)}
\biggl|_{j,J=0}=\frac{1}{Z^{(n)}}\frac{\delta^2 Z^{(n)}}{\delta j_\mu(x)\delta
j_\nu(y)}\biggl|_{j,J=0}=-\alpha^2 r_0^2<B^{(0)\mu}(x)B^{(0)\nu}(y)>
\ee
and
\be
\frac{1}{Z^{(n)}_M}\frac{\delta^2 Z^{(n)}_M}{\delta J_\mu(x)\delta J_\nu(y)}
\biggl|_{j,J=0}=\frac{1}{Z^{(n)}_D}\frac{\delta^2 Z^{(n)}_D}{\delta J_\mu(x)
\delta J_\nu(y)}\biggl|_{j,J=0}=-\alpha^2<F^\mu(A)(x)F^\nu(A)(x)>_D
\ee
The master theory allows writing
\be
\frac{1}{Z^{(n)}_M}\frac{\delta^2 Z^{(n)}_M}{\delta j_\mu(x)\delta j_\nu(y)}
\biggl|_{j,J=0}=\frac{1}{Z^{(n)}_D}\frac{\delta^2 Z^{(n)}_D}{\delta j_\mu(x)
\delta j_\nu(y)}\biggl|_{j,J=0}=-\alpha^2<F^\mu(A)(x)F^\nu(A)(y)>_D-
i\frac{\alpha^2r_0}{2}\delta^{\mu\nu}\delta(x-y)
\ee
and
\be
\frac{1}{Z^{(n)}_M}\frac{\delta^2 Z^{(n)}_M}{\delta J_\mu(x)\delta J_\nu(y)}
\biggl|_{j,J=0}=\frac{1}{Z^{(n)}}\frac{\delta^2 Z^{(n)}}{\delta J_\mu(x)\delta
J_\nu(y)}\biggl|_{j,J=0}=-\alpha^2<F^\mu(B)(x)F^\nu(B)(y)>+i\frac{\alpha^2}{2}
O^{\mu\nu}\delta(x-y)
\ee
where the operator $O^{\mu\nu}$ is defined according to
\be
F^\mu(A)=O^{\mu\nu}A^{(0)}_\nu
\ee
We use these results to write
\be\label{q1}
<B^{(0)}_\mu(x)B^{(0)}_\nu(y)>=<F_\mu(A)(x)F_\nu(A)(y)>_D+
i\frac{r_0}{2}\delta_{\mu\nu}\delta(x-y)
\ee
and
\be\label{q2}
<F^\mu(B)(x)F^\nu(B)(y)>-i\frac12 O^{\mu\nu}\delta(x-y)=
<F^\mu(A)(x)F^\nu(A)(y)>_D
\ee

Expressions (\ref{q1}) and (\ref{q2}) show that the correponding Green
functions are equivalent, apart from contact terms. They show that the
duality that we have presented in Sec.~(III.A) above is preserved at the
quantum level.

We now examine the specific case considered in Sec.~{\ref{sec:exe}}, which
involves duality between the MCS-Proca theory (\ref{mcsp}) and the generalized
MCS-Podolsky model (\ref{mcspo}). In this case the results (\ref{q1})
and (\ref{q2}) become
\be
<B_\mu(x)B_\nu(y)>=<F_\mu(A)(x)F_\nu(A)(y)>_D+i\frac{m^2}{4}\delta_{\mu\nu}
\delta(x-y)
\ee 
and
\be
<F_\mu(B)(x)F_\nu(B)(y)>-\frac{i}{2}\left(-\frac{m}{2}\varepsilon_{\mu\nu
\lambda}\partial^\lambda+\frac{a}{2}\delta_{\mu\nu}\square-\frac{a}{2}
\partial_\mu\partial_\nu\right)\delta(x-y)=<F_\mu(A)(x)F_\nu(A)(y)>_D
\ee
where we have re-inserted the parameters used in Sec.~{\ref{sec:exe}}.
We see that for $a=0$ one recovers the result of Ref.~{\cite{brr}}, which deals
with quantum duality for the CS and MCS models.

%%%%%%%%%%%%%%%%%%%%%%%%%%%%%%%%%%%%%%%%%%%%%%%%%%%%%%%%%
\section{The case $n\to\infty$}
\label{sec:og}

The success of the former investigations has lead us to think on extending
the case containing a finite number of terms to the case where an infinity
sequence of terms is considered. Our interest relies on the case
where the infinity sequence of terms adds to give elementary functions.
We see that the form of ${\cal L}^{(n)}$ in (\ref{gt})
suggests this particular generalization, to the case where
$n\to\infty$. Such generalization can be written
in terms of a smooth function, depending on the specific values of the real
parameters $r_i$ that we have introduced to define the model. We
further explore this possibility introducing the model
 \be
  {\cal L}=A_\mu[F({\cal O})]^\mu_{\;\,\nu}A^\nu
 \ee
where $F$ is a smooth function, and ${\cal O}$ is the operator
 \be
{\cal O}^{\mu\nu}\equiv
-\varepsilon^{\mu\nu\lambda}\partial_{\lambda}
 \ee
The model is defined in terms of the expansion of the smooth
function, in the form \be {\cal L}=A_\mu\sum_{n=0}^{\infty}
C_n[{\cal O}^n]^\mu_{\;\,\nu}A^\nu
 \ee
where $C_n$ are given by
 \be C_n=\frac1{n!}\frac{d^n
F(x)}{dx^n}\Bigl|_{x=0}
 \ee
and $[{\cal O}^0]^\mu_{\;\nu}=\delta^\mu_{\;\,\nu}$,
$[{\cal O}^1]^\mu_{\;\nu}=-\varepsilon^{\mu}_{\;\,\nu\lambda}
\partial^\lambda$ and
$[{\cal O}^2]^\mu_{\;\nu}=[{\cal O}^1]^\mu_{\;\lambda}
[{\cal O}^1]^\lambda_{\;\nu}$,
and so forth. We require that $F(0)\neq0$,
which implies that $C_0\neq0$. This means that the above model starts
with the Proca-like term $A_\mu A^\mu$, which prevents the
presence of gauge invariance. Thus, we use the gauge embedding
procedure to obtain the dual model. It has the form
 \be
{\cal L}_{D}=A_\mu\sum^\infty_{n=0}C_n\left[{\cal O}^n
\right]^{\mu}_{\,\,\nu}A^\nu-\frac1{C_0}A_\mu\sum^\infty_{n=0}
\sum^\infty_{m=0}C_n C_m\left[{\cal O}^{n+m}\right]^{\mu}_{\,\,\nu} A^\nu
 \ee
or, formally,
\be
{\cal L}_{D}=A_\mu\left[F({\cal O})-\frac1{F(0)}\,[F({\cal O})]^2
\right]^{\mu}_{\,\,\nu}A^\nu
\ee

We illustrate this general result with the example
\be
{\cal L}=A_\mu\left(\frac{1}{1+{\cal O}}\right)^\mu_{\;\nu}A^\nu
\ee
In this case the dual theory reads
\be
{\cal L}_D=A_\mu\left(\frac{\cal O}{(1+{\cal O})^2}\right)^\mu_{\;\nu}A^\nu
\ee

%%%%%%%%%%%%%%%%%%%%%%%%%%%%%%%%%%%%%%%%%%
\section{Adding Fermions}
\label{sec:mat}

The study of fermions is motivated by the possiblity of extending the
present duality procedure to more realistic models, which should necessarily
contain fermionic matter fields to describe the matter contents of any
realistic model.

We add fermions with the modification
 \be\label{ft}
{\wt{\cal L}}^{(n)}={\cal L}^{(n)}+{\cal L}_{I}+{\cal L}_f
 \ee
where
 \be
{\cal L}_f={\bar\psi}(i{/\!\!\!}\partial-M){\psi}
 \ee
describes free fermions, with $M$ being the dimensionless fermion mass
parameter. To identify how the fermionic field interacts with the fields
$A^{(n)}_\mu$ we introduce the non-minimal coupling
 \be
\partial_\mu\to D_\mu=\partial_\mu+i\sum^n_{i=0}e_i A_\mu^{(i)}
 \ee
where $e_i$ are (dimensionless) coupling constants -- we notice that
the condition $e_i=0,\, i\neq0$ leads to minimal coupling with the
fermionic matter. With the above generic coupling, the interactions
terms in ${\cal L}_I$ are given by
 \be
{\cal L}_{I}=-\sum^n_{i=0}e_i A^{(0)\mu}J_{\mu}^{(i)}
 \ee
where we have defined the general current
 \be
J_{\mu}^{(i)}\equiv\varepsilon_{\mu\nu\lambda}\partial^\nu
J^{(i-1)\lambda}
 \ee
with
 \be
J^{(0)}_\mu\equiv j_\mu=\bar{\psi}\gamma_\mu\psi
 \ee
To write the above expressions we have used the identity
 \be
A^{(i)}_\mu J^{(j)\mu}=A^{(i-1)}_\mu J^{(j+1)\mu}+
\varepsilon^{\mu\nu\lambda}\partial_\mu
\bigl[A^{(i-1)}_\nu J^{(j)}_\lambda\bigr]
 \ee

We now search for the dual theory. The procedure follows as in the former
case. The Euler vector is modified by the presence of interactions; it changes
to
 \be
{\wt K}_\mu=K_\mu-\sum^n_{i=0}e_i J^{(i)}_\mu
 \ee
We introduce an auxiliary field ${\wt a}_\mu$, and we impose that
$\delta{\wt a}_\mu=\delta A^{(0)}_\mu=\delta\Lambda$. The Lagrange density
varies according to $\delta{\cal L}^{(1)}=-{\wt a}_\mu\delta
{\wt K}^\mu$, since $\delta J^{(n)}_\mu=0$, as one can verify
straightforwardly. Like in the former case, the procedure requires
another iteration. The final result is
 \be
{\wt{\cal L}}^{(n)}_{D}={\wt{\cal L}}^{(n)}-\frac{1}{4r_0}{\wt
K}^\mu{\wt K}_{\mu}
 \ee
or, explicitly,
 \be\label{dft}
{\wt{\cal L}}^{(n)}_{D}=-\sum_{i=1}^{2n}\left(r'_i A^{(0)}_\mu
A^{(i)\mu}-s'_i A^{(0)}_\mu J^{(i)\mu}\right)-\frac14\sum_{i=0}^{2n}e'_i
J_\mu^{(0)}J^{(i)\mu}+{\cal L}_f
 \ee
where we have set
\be
s_i' = \begin{cases}
-\dfrac{1}{r_0}\displaystyle\sum_{l=1}^i r_l \,e_{i-l}&
\text{for $1\leqslant i \leqslant n$}\\ \\
-\dfrac{1}{r_0}\displaystyle\sum_{l=i-n}^{n} r_l \,e_{i-l} &
\text{for $n<i\leqslant2n$}\end{cases}
\ee
and
\be
e_i' = \begin{cases}
\dfrac{1}{r_0}\displaystyle\sum_{l=0}^i e_l \,e_{i-l}&
\text{for $0\leqslant i \leqslant n$}\\ \\
\dfrac{1}{r_0}\displaystyle\sum_{l=i-n}^{n} e_l \,e_{i-l} &
\text{for $n<i\leqslant2n$}\end{cases}
\ee

In this result, we notice the presence of Thirring-like
interactions \cite{ith}, which are fundamental to maintain the contents
of the fermionic sector unchanged \cite{go}. To see this we examine the
dynamics of the fermionic sectors in both theories. The fermionic equation
of motion for the first theory is
 \be\label{ef}
(i{/\!\!\!\partial}-M)\psi=\sum_{i=0}^n e_iA^{(i)}_\mu
\gamma^\mu\psi
 \ee
To eliminate the gauge field, we notice that
 \be
\sum_{i=0}^n r_i\,A^{(i)}_\mu=\frac12 \sum_{i=0}^n e_i\,J^{(i)}_\mu
 \ee
We restrict the coupling constants to obey $e_i=\alpha r_i$ to get
\be\label{ef1}
(i{/\!\!\!\partial}-M)\psi=\frac12\alpha\sum_{i=0}^n e_i\,J^{(i)}_\mu
\gamma^\mu\psi
 \ee

Analogously, the fermionic equation for the dual theory is given by
 \be
(i{/\!\!\!\partial}-M)\psi=\sum_{i=1}^{2n}s'_iA^{(i)}_\mu
\gamma^\mu\psi+\frac12 \sum_{i=1}^{2n}e'_iJ^{(i)}_\mu
\gamma^\mu\psi
 \ee
We find the equation of motion for the gauge field in the dual model as
 \be
\sum_{i=1}^{2n}r'_i A^{(i)}_\mu=\frac12
\alpha\sum_{i=1}^{2n}s'_iJ^{(i)}_\mu
 \ee
The restriction $e_i=\alpha r_i$ implies that $s'_i=\alpha r'_i$, and so
we can write
 \be
(i{/\!\!\!\partial}-M)\psi=\frac12\alpha\sum_{i=0}^{n}e_i\,J^{(i)}_\mu
\gamma^\mu\psi
 \ee
which is Eq.~(\ref{ef1}). This result shows that the fermionic sector does
not change when one goes from (\ref{ft}) to the dual theory (\ref{dft}).

We can also add fermions in the case $n\to\infty$. To illustrate this
possibility we consider the model
\be
{\wt {\cal L}}=A_\mu[F({\cal O})]^\mu_{\;\;\nu}A^\nu-
A_\mu[G({\cal O})]^\mu_{\;\;\nu}J^\nu+{\cal L}_f
\ee
where $G$ is a smooth function similar to $F$. We use the gauge embedding
procedure to obtain the dual model
\be
 {\wt {\cal L}}_D=A_\mu\left[F({\cal O})-
\frac1{F(0)}F^2({\cal O})\right]^\mu_{\;\;\nu}A^\nu-
A_\mu\left[G({\cal O})-\frac1{F(0)}F({\cal O})G({\cal O})\right]^\mu_{\;\;\nu}
J^\nu-\frac14 J_\mu\left[\frac1{F(0)}G^2({\cal O})\right]^\mu_{\;\,\nu}J^\nu+
{\cal L}_f
\ee
We notice the presence of the generalized Thirring-like term in the dual
theory.

%%%%%%%%%%%%%%%%%%%%%%%%%%%%%%%%%%%%%%%%%%%%%%%%%%%%%%%%%
\section{Nonlinear interactions}
\label{sec:nli}

We now return to the Born-Infeld \cite{bi} generalization, which is different
from the Proca and Podolsky generalizations.  The main ingredient now is
nonlinearity, and so we further explore the duality procedure introduced above
mixing higher-order derivatives and nonlinearity. Evidently, there are
several different possibilities of including nonlinearity in the model
introduced in the former Sec.~{\ref{sec:gen}}, but here we consider the case
\be
{\cal L}^{(n)}_{NL}=g(A^{(0)}_\mu A^{(0)\mu})+
\sum^{n}_{i=1}r_i A_\mu^{(0)}A^{(i)\mu}
\ee
where $g(x)$ is nonlinear in $x=A^{(0)}_\mu A^{(0)\mu}$. This case does
not include nonlinear interactions that involve derivative of the basic
field $A_\mu=A_\mu^{(0)}$. The equation of motion is
 \be
A^{(0)}_\mu=-\frac1{g^\prime(x)}\sum_{i=1}^{n}r_i A_\mu^{(i)}
 \ee
where $g^{\prime}(x)=dg/dx$. This equation allows writing
\be
\partial^\mu A^{(0)}_\mu=-\sum_{i=1}^{n}r_i A_\mu^{(i)}\;
\partial^\mu\left(\frac1{g^\prime(x)}\right)
 \ee
which shows that there are longitudinal mode propagating, due to the presence
of the nonlinear interaction. We notice that in the linear case [$g(x)=x$]
we have $g^\prime=1$, which leaves no room for propagation of longitudinal
modes.

We treat the presence of nonlinearity invoking the trick used in
Ref.~{\cite{e3}}, and then further explored in \cite{e4,mnrw}. The key point
here is to remove the nonlinearity under the expense of 
introducing another field, an auxiliary scalar field $\phi$. We implement
this possibility with the change
\be
g(A^{(0)}_\mu A^{(0)\mu})\to f(\phi)+\frac1{\phi} A^{(0)}_\mu A^{(0)\mu}
\ee
We follow \cite{mnrw} to show that
\be\label{fphi}
f(\phi)=\int^\phi d\chi \,\frac1{\chi^2}\,
g^{\prime-1}\left(\frac1{\chi}\right)
\ee
The model is modified to
\be
{\cal L}^{(n)}_\phi=f(\phi)+\frac1{\phi}A^{(0)}_\mu A^{(0)\mu}+
\sum^{n}_{i=1}r_i A_\mu^{(0)}A^{(i)\mu}
\ee
In this case the Euler vector is given by
\be
K_\mu=\frac2{\phi}A^{(0)}_\mu+
2\,\sum^{n}_{i=1}r_i A^{(i)}_\mu
\ee
The gauge embedding method allows writing
\be
{\cal L}^{(n)}_D={\cal L}^{(n)}_\phi-\frac14\phi K_\mu K^\mu
\ee
and so
\be\label{nld}
{\cal L}^{(n)}_D=f(\phi)-\sum^{n}_{i=1}r_i A_\mu^{(0)}A^{(i)\mu}-
\phi\sum^{2n}_{i=2}{\tilde r}_i A_\mu^{(0)}A^{(i)\mu}
\ee
where ${\tilde r}_i$ is given by
\be
{\tilde r}_i = \begin{cases}
\;\;\displaystyle\sum_{l=1}^{i-1}r_l \,r_{i-l}&
\text{for $2\leqslant i \leqslant n$}\\ \\
\;\;\displaystyle\sum_{l=i-n}^{n}r_l \,r_{i-l} &
\text{for $n<i\leqslant2n$}\end{cases}
\ee
We notice that in (\ref{nld}) the nonlinear behavior involve all the terms,
unless the Chern-Simons one. This fact is clearer when we eliminate
the auxiliary field $\phi$ from the model, which is formally given
by
\be\label{phi}
\phi={f^{\prime}}^{-1}
\left(\sum^{2n}_{i=2}{\tilde r}_i A_\mu^{(0)}A^{(i)\mu}\right)
\ee

We illustrate the above investigations with the model
\be\label{nlbi}
{\cal L}^{(n)}_{BI}=\beta^2\sqrt{1+
\frac{2r_0}{\beta^2}A_\mu^{(0)}A^{(0)\mu}}+
\sum_{i=1}^n r_i A_\mu^{(0)}A^{(i)\mu}
\ee
where the nonlinear contribution is of the Born-Infeld type.
We notice that the limit $\beta\to\infty$ restores the original model
(\ref{gt}). In this case, the function $f(\phi)$ given by Eq.~(\ref{fphi})
has the form
\be
f(\phi)=\frac12\beta^2\left(r_0\phi+\frac1{r_0\phi}\right)
\ee
We use Eq.~(\ref{phi}) to obtain
\be
\phi=\frac1{r_0}\Biggl{/}{\sqrt{1-\frac2{r_0\beta^2}
\sum^{2n}_{i=2}{\tilde r}_i A_\mu^{(0)}A^{(i)\mu}}}
\ee
The gauge embedding procedure given above allows writing
\be
{\cal L}^{(n)}_{BID}=\beta^2\sqrt{1-
\frac{2}{r_0\beta^2}\sum_{i=2}^{2n}{\tilde r}_i A_\mu^{(0)}A^{(i)\mu}}-
\sum_{i=1}^n r_i A_\mu^{(0)}A^{(i)\mu}
\ee
We see that in the limit $\beta\to\infty$ the above result leads to
(\ref{dgt}), the dual of the model (\ref{gt}), as expected. We also notice
that in the case $n=1$ the model (\ref{nlbi}) reproduces the
Born-Infeld-Chern-Simons model investigated in Ref.~{\cite{e3}},
and this shows that the model (\ref{nlbi}) is a generalization of the model
investigated in \cite{e3}.

%%%%%%%%%%%%%%%%%%%%%%%%%%%%%%%%%%%%%%%%%
\section{Comments and conclusions}
\label{sec:cc}

In the present work we have investigated duality symmetry in generalized
Field Theory models, involving higher-order derivative of the basic field
$A_\mu=A^{(0)}_\mu$ through the recursive relation
$A^{(n)}_\mu=\varepsilon_{\mu}^{\;\;\nu\lambda}\partial_\nu A^{(n-1)}_\lambda$.
The investigations started in Sec.~{\ref{sec:exe}} and in Sec.~{\ref{sec:gen}},
and there we have generalized the self-dual model to include several
higher-order derivative terms, and we have obtained the dual
theory. We have also proposed a master model,
from which one gets both the model and its dual
partner. In this generalization, the presence of the CS and CS-like terms
imposes the restriction that the Minkowsky space has $(2,1)$ spacetime
dimensions. However, we can eliminate all the CS and CS-like terms
with the restriction $r_i=0$ for $i$ odd; in this case
our results are valid in arbitrary $(d,1)$ space-time dimensions.
In Sec.~{\ref{sec:gen}} we have also examined the implications of duality
at the quantum level. These results are obtained for ${\cal L}^n$, for $n$
integer, finite, and in Sec.~{\ref{sec:og}} we have further extended our
results, considering the limit $n\to\infty$, which leads to the case involving
non-polynomial functions.

In the sequel, in Sec.~{\ref{sec:mat}} we have added fermions to the system,
and there we have included fermions in models involving $A^{(n)}_\mu$,
in the case where $n$ is finite, and also for $n\to\infty$. In
Sec.~{\ref{sec:nli}} we have investigated the more general case,
where nonlinear contributions involving the basic field
$A^{(0)}_\mu$ is present. As we have shown, in this
case we first circumvent nonlinearity at the expense of introducing an
auxiliary field, and then we proceed as before, to get to the dual model.
After getting to the dual model, we then eliminate the auxiliary field to
obtain the dual model in terms of the original field $A^{(0)}_\mu$
and the accompanying derivatives. 

We notice that all the generalizations we have investigated involve the
Abelian vector field $A_\mu$. Thus, a natural and direct issue concerns
the use of non-Abelian fields, and their possible generalizations,
along the lines of the non-Abelian SD model, and the
Yang-Mills-Chern-Simons model \cite{e1}. Also, it is of
interest to investigate if the presence of bosonic matter changes
the duality scenario that we have presented in Sec.~{\ref{sec:mat}}.
Another point concerns duality in models of the B$\wedge$F type, and their
extensions to include higher-order derivatives. These and other related issues
are under consideration, and we hope to report on them in the near future.
Another point concerns models which include fermions. As one knows, if there
is no fermionic self-interaction, and if one integrates on the fermions,
the remaining action will necessarily contain higher-order derivative terms,
thus giving another compelling motivation to investigate models involving
higher-order derivatives. We hope to report on these and other related
issues in another work.

We would like to thank CAPES, CNPq, PROCAD and PRONEX for partial support.

%%%%%%%%%%%%%%%%%%%%%%%%%%%%%%%%%%%%%%%%%%%%

\end{document}